\documentclass[10pt,twocolumn,letterpaper]{article}

\usepackage{cvpr}              
\usepackage{multirow}

\usepackage{booktabs}
\usepackage{makecell}
\usepackage{array}
\usepackage{graphicx}
\usepackage{threeparttable}

\usepackage{multirow}

\usepackage{comment}
\usepackage{cancel}
\usepackage{wasysym} 

\newcommand{\state}{\boldsymbol{s}}

\newcommand{\x}{\boldsymbol{x}}

\newcommand{\n}{\boldsymbol{n}}
\newcommand{\valpha}{\boldsymbol{\alpha}}
\newcommand{\vbeta}{\boldsymbol{\beta}}
\newcommand{\pol}{\hat{\boldsymbol{e}}} 

\usepackage{pifont}
\definecolor{cvprblue}{rgb}{0.21,0.49,0.74}
\usepackage[pagebackref,breaklinks,colorlinks,allcolors=cvprblue]{hyperref}

\def\paperID{18678} 
\def\confName{CVPR}
\def\confYear{2026}

\title{MetaSpectra+: A Compact Broadband Metasurface Camera\\ for Snapshot Hyperspectral+ Imaging }

\author{Yuxuan Liu$^{*}$, Wei Xu$^{*}$, and Qi Guo\\
Elmore Family School of Electrical and Computer Engineering, Purdue University\\
{\tt\small \{liu3910,weixu,qiguo\}@purdue.edu}\\[0.3em]
{\small $^{*}$Equal contribution}
}

\begin{document}
\maketitle

\begin{abstract}
We present MetaSpectra+, a compact multifunctional camera that supports two operating modes: (1) snapshot HDR + hyperspectral or (2) snapshot polarization + hyperspectral imaging. It utilizes a novel metasurface-refractive assembly that splits the incident beam into multiple channels and independently controls each channel’s dispersion, exposure, and polarization. Unlike prior multifunctional metasurface imagers restricted to narrow (10--100 nm) bands, MetaSpectra+ operates over nearly the entire visible spectrum (250 nm). Relative to snapshot hyperspectral imagers, it achieves the shortest total track length and the highest reconstruction accuracy on benchmark datasets. The demonstrated prototype reconstructs high-quality hyperspectral datacubes and either an HDR image or two orthogonal polarization channels from a single snapshot.
\end{abstract}

\section{Introduction}
\label{sec:intro}

The emergence of multifunctional metasurfaces has enabled the simultaneous acquisition of multiple imaging modalities within a compact, monocular form factor~\cite{khorasaninejad2016multispectral, guo2019compact, shi2020continuous, hazineh2023polarization,brookshire2024metahdr, liu2025metah2}. However, such systems remain fundamentally limited by severe chromatic aberrations, which confine their operational bandwidth to a single wavelength or a narrow spectral range (typically 10--100~nm). This constraint significantly hinders their versatility and practical deployment. Therefore, expanding the usable bandwidth of multifunctional metasurface imagers is crucial to unlocking their integration into a wider range of applications~\cite{jiang2018gphenovision, lahat2015multimodal}, ultimately enabling compact, snapshot, multi-modal imaging with accurate cross-modal registration.

\begin{figure}
    \centering
    \includegraphics[width=1.0\linewidth]{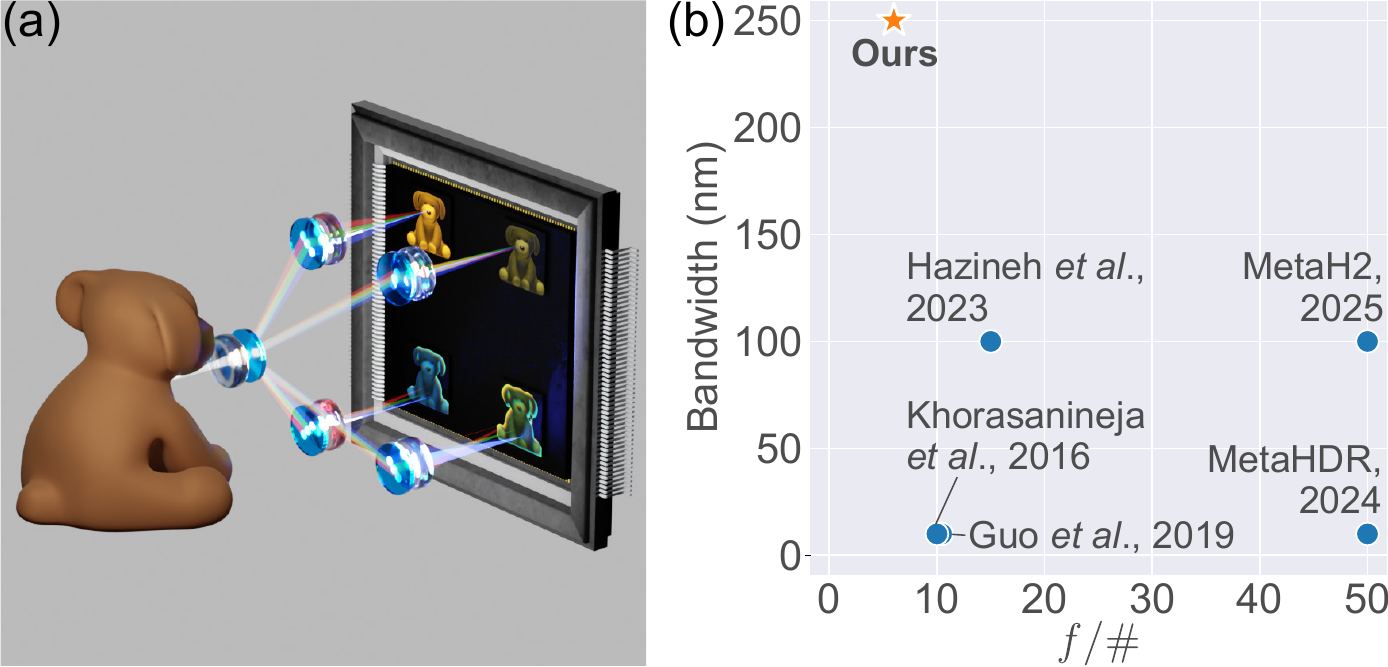}
    \caption{Overview. (a) MetaSpectra+ employs a compact, hybrid optical assembly that integrates refractive lenses with metasurfaces (blue), forming multiple images in a single shot, each engineered with distinct dispersion, exposure, or polarization. The system reconstructs a hyperspectral datacube together with either an HDR image or two polarization channels from a snapshot capture. (b) Compared with previous multifunctional metasurface systems~\cite{khorasaninejad2016multispectral, guo2019compact, hazineh2023polarization, brookshire2024metahdr, liu2025metah2}, the hybrid optical design of MetaSpectra+ supports a significantly broader operating bandwidth and a lower F-number, while being comparably compact.}
    \label{fig:teaser-figure}
\end{figure}

We propose \textit{MetaSpectra+}, a broadband, metasurface-refractive, multifunctional imaging system. As illustrated in \cref{fig:teaser-figure}a, the system integrates a novel two-layer metasurface assembly: the first layer splits and deflects the incident beam into multiple optical channels, while the second layer partially or completely compensates for this deflection to either introduce a controlled amount of or eliminate dispersion in the captured images. Owing to this dispersion control, it supports a substantially broader operational bandwidth (250 nm) than previous multifunctional metasurface systems. In addition, MetaSpectra+ decouples image formation from beam splitting, utilizing refractive optics and metasurfaces to perform each, respectively. This separation enables the system to operate at a significantly lower F-number while preserving a compact form factor, outperforming prior designs that rely on a single multifunctional metasurface to simultaneously perform beam splitting and image formation~\cite{brookshire2024metahdr, liu2025metah2}. See \cref{fig:teaser-figure}b.

In this paper, we demonstrate a MetaSpectra+ prototype capable of performing hyperspectral and HDR imaging simultaneously over a wavelength band from 450~nm to 700~nm. As illustrated in \cref{fig:teaser-figure}a, the system captures two achromatic images and two chromatic images with orthogonal dispersions. The achromatic image pair forms an exposure bracket~\cite{debevec2023recovering}, enabling the reconstruction of a high-dynamic-range (HDR) image of the scene. Meanwhile, the four captured images collectively constitute a computational tomographic imaging spectroscopy (CTIS) configuration~\cite{descour1995computed}, allowing hyperspectral datacube reconstruction. We further show that the achromatic images can be reconfigured to capture two orthogonal polarizations, enabling simultaneous hyperspectral and polarization imaging.



According to our simulation and real-world analyses, MetaSpectra+ achieves the highest hyperspectral reconstruction accuracy on benchmark datasets compared to the latest snapshot hyperspectral imaging systems. Furthermore, the prototype demonstrates snapshot acquisition of high-quality HDR images and hyperspectral data cubes, achieving 11~dB increase in dynamic range on real captured data compared to without the exposure bracketing. 

The contribution of this paper is as follows:
\begin{enumerate}
    \item A new metasurface-refractive optical paradigm for broadband multifunctional imaging;
    \item A MetaSpectra+ prototype for simultaneous hyperspectral reconstruction and HDR/polarization imaging;
    \item A comprehensive analysis that demonstrates the state-of-the-art performance of the proposed system.
\end{enumerate}
All data and code of this work can be accessed at \href{https://meta-imaging.qiguo.org}{https://meta-imaging.qiguo.org}.














\section{Related Work}
\label{sec:related}

\subsection{Snapshot Hyperspectral Imaging Systems}

Snapshot hyperspectral imaging seeks to recover a three-dimensional hyperspectral datacube from one or a few simultaneously captured two-dimensional sensor measurements~\cite{thomas2025trends, hagen2012snapshot, li2021spectral}. This dimensionality gap requires the optical system to either sample the hyperspectral datacube~\cite{dwight2017lenslet, ji2023spatial, zhu2025snapshot, yu2021batch}; encode the hyperspectral information through engineered optical sensitivities~\cite{zhang2025lensless, baek2017compact, arguello2021shift}; or combine both strategies~\cite{gao2009compact, kim2025snapshot}. Meanwhile, snapshot approaches drastically reduce the acquisition time and latency compared to traditional multi-shot hyperspectral imaging systems~\cite{henriksen2022yourself, aydin2025spectrum, foley2025spectral, fowler2014compressive}.
Here, we review representative solutions for snapshot hyperspectral imaging with respect to hardware compactness, reconstruction fidelity, light efficiency, and other system-level trade-offs.

\paragraph{Sampling-based Approaches} acquire a sparsely sampled subset of the hyperspectral datacube in either the spatial or spectral domain, and reconstruct the full datacube by imposing priors on its spatial–spectral structure~\cite{chakrabarti2011statistics}.
Spatial-domain sampling is typically implemented using coded apertures~\cite{wagadarikar2008single}, spatial light modulators (SLMs)~\cite{saragadam2019krism, saragadam2021sassi}, lenslet arrays~\cite{dwight2017lenslet, ji2023spatial, deng2025compact}, or fiber bundles~\cite{french2018snapshot, bedard2012snapshot}, each offering different adaptability and resolution. These platforms generally require relay optics, leading to bulkier form factors.
Spectral-domain sampling is achieved using spectral filter arrays (SFAs) with standard~\cite{zhao2020hierarchical} or custom~\cite{saragadam2020programmable, zhu2025snapshot} filter mosaics, or through dispersive spectral filters~\cite{harvey2003high, murakami2012hybrid, hua2022ultra, mcclung2020snapshot} for wavelength-dependent beam splitting. These designs are typically compact but incur high fabrication cost. Under a fixed number of sensor pixels, both approaches trade off spatial and spectral resolutions to meet the requirements of target applications. 









\paragraph{Encoding-based Approaches} embed spectral information into the spatial domain through engineered, wavelength-dependent point spread functions (PSFs). Compared to sampling-based methods, these systems are typically more compact and light-efficient, but require more sophisticated computational reconstruction.
Spectral encoding has been realized using spatial light modulators (SLMs)~\cite{zhang2025lensless, zhang2025tunable}, prisms~\cite{baek2017compact}, diffusers~\cite{monakhova2020spectral}, diffractive optical elements (DOEs)~\cite{arguello2021shift, li2022quantization, dun2020learned, jeon2019compact, shao2025optimization, kim2025snapshot, shi2024learned, shi2024split}, and gratings~\cite{wu2023ctis, douarre2021ctis, kudenov2012faceted}. Active illumination has also been explored to encode spectral information via time multiplexing~\cite{verma2024chromaflash} or controlled dispersion~\cite{shin2024dispersed}.
These systems leverage recent advances in learning-based image reconstruction~\cite{ho2020denoising, dong2021dwdn, hazineh2024grayscale} to decode the encoded measurements, enabling high-quality spatial–spectral recovery.























\subsection{Multifunctional Imaging}

Many applications, such as agricultural phenotyping~\cite{jiang2018gphenovision} and forensics~\cite{lahat2015multimodal}, benefit from accurately registered multimodal visual data that includes hyperspectral information. Conventional multimodal fusion relies on sophisticated cross-modal registration algorithms~\cite{velesaca2024multimodal} to achieve satisfactory alignment across sensing modalities. To circumvent this challenge, several monocular systems have been developed to jointly encode spectral information together with additional scene attributes—such as depth~\cite{deng2025compact, baek2021single} or spatial frequency~\cite{urban2021multimodal}—within a single measurement for fused reconstruction.

The rapid progress of metasurface optics in recent years has enabled compact and highly integrated multifunctional imagers. A single metasurface can split the incident beam into multiple optical channels based on wavelength~\cite{hua2022ultra}, polarization~\cite{hazineh2023polarization, khorasaninejad2016multispectral}, or incident angle~\cite{shi2020continuous}, and apply independent optical modulation to each channel to realize distinct focal settings~\cite{guo2019compact}, PSF designs~\cite{hazineh2023polarization}, dynamic ranges~\cite{brookshire2024metahdr}, or dispersion directions~\cite{liu2025metah2}. Since all measurements are captured from a single viewpoint, the resulting modalities are inherently registered.

However, the strong chromatic dispersion of metasurfaces fundamentally limits their achromatic performance. Despite recent advances in broadband metasurfaces for imaging~\cite{froch2025beating, chakravarthula2023thin, tseng2021neural, sun2025collaborative}, most multifunctional metasurfaces operate over only a narrow spectral band (10--100~nm)~\cite{khorasaninejad2016multispectral, brookshire2024metahdr, liu2025metah2}, restricting their practical utility, especially for hyperspectral imaging. To our knowledge, no hyperspectral imagers using multifunctional metasurface has demonstrated operations across the full visible spectrum.









\section{System Overview}
\label{sec:opticaldesign}

\subsection{Optical Model}

\begin{figure}
    \centering
    \includegraphics[width=\linewidth]{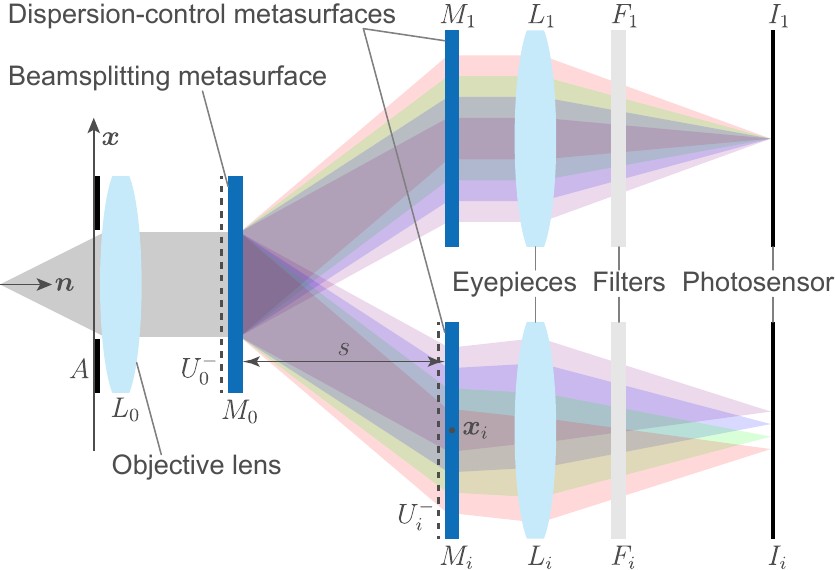}
    \caption{MetaSpectra+ optical design. The optical assembly simultaneously captures multiple sub-images $I_{1:V}$ of the same scene from broadband illumination, each with independently controlled dispersion, exposure, and polarization sensitivity, using a compact two-layer metasurface design.}
    \label{fig:image-formation-model}
\end{figure}

\Cref{fig:image-formation-model} illustrates the proposed optical assembly, which integrates refractive optics, metasurfaces, and optical filters into a compact layered configuration. An objective lens with a field stop $A(\x)$ collimates the light from an incoherent point source within the depth of field. The collimated beam is then divided and deflected by the beamsplitting metasurface $M_0$ into $V$ subsequent optical channels. 

Consider a broadband collimated beam propagating along direction $\n = (n_x, n_y, n_z)^\top$ before reaching $M_0$. Its planar wavefront can be expressed as:
\begin{align}
    U_0^-(\x, \lambda, \n) = \exp\left(j\frac{2\pi}{\lambda}\n_{\perp}\cdot\x\right), 
\end{align}
where $\n_{\perp}$ represent the transverse component of the propagating direction $\n$, $\n_{\perp} = (n_x, n_y)^\top$, and $\lambda$ denotes the wavelength.

\paragraph{Beamsplitting Metasurface.}
As shown in \cref{fig:image-formation-model}, the beamsplitting metasurface $M_0$ divides and angularly deflects the incident collimated beam into $V$ distinct optical channels. 
To steer the incident wavefront toward the $i$th channel, a linear phase delay needs to be imparted at the design wavelength $\lambda_c$:
\begin{align}
    M_{0,i}(\x, \lambda_c) = 
    \exp\!\left(j\frac{2\pi}{\lambda_c}\valpha_i\cdot\x\right),
    \quad i = 1,\dots,V.
\end{align}
Because $M_{0,i}(\x, \lambda_c)$ is a periodic function, its modulation for other wavelengths $\lambda$ can be expressed as a Fourier series:
\begin{align}
    M_{0,i}(\x, \lambda) = 
    \sum_{n=-\infty}^{\infty} 
    a_n(\lambda)\,
    \exp\!\left(j\frac{2\pi n}{\lambda_c}\valpha_i\cdot\x\right),
\end{align}
where $a_n(\lambda)$ denotes the diffraction efficiency of the $n$th order arising from chromatic dispersion. 
Empirically, only the zeroth and first orders ($a_0$, $a_1$) contribute significantly, while higher orders are negligible. 
Moreover, the field stop of the subsequent metasurface layer naturally blocks the zeroth-order diffraction term $a_0$. 
Therefore, the effective modulation of $M_0$ over the visible spectrum can be approximated as:
\begin{align}
    M_{0,i}(\x, \lambda) \approx 
    a_1(\lambda)\, M_{0,i}(\x, \lambda_c).
    \label{eq:deflect1}
\end{align}
To realize \emph{simultaneous beamsplitting}, the overall phase profile of $M_0$ at $\lambda_c$ is constructed as a random interleaving of the individual sub-profiles $M_{0,k}(\x, \lambda_c)$, following the approach of Brookshire \etal~\cite{brookshire2024metahdr}:
\begin{equation}
    \begin{aligned}
    M_0(\x, \lambda_c) = 
    M_{0,k}(\x, \lambda_c),\quad k\sim\text{Multinomial}(1/V),
    \label{eq:interleave}
\end{aligned}
\end{equation}
where the overall phase profile is sampled from $M_{0,1:V}$ following an equal-weight multinomial distribution. This randomized interleaving allows the metasurface $M_0$ to simultaneously form $V$ beams at large deflection angles while suppressing residual diffraction artifacts introduced by the spatial multiplexing of optical modulation profiles $M_{0,1:V}$.

\paragraph{Dispersion-Control Metasurface.}
According to \cref{eq:deflect1}, the collimated beam arriving at the $i$th metasurface $M_i$ ($i=1,\dots,V$) exhibits wavelength-dependent angular dispersion. The incident wavefront at $M_i$ can be written as
\begin{equation}
    \begin{aligned}
    U_i^{-}&(\x, \lambda, \n) = \\
    &A(\x, \lambda) a_1(\lambda) \exp{\left(j\frac{2\pi}{\lambda}\left(\left[\frac{\lambda\valpha_i}{\lambda_c} + \n_{\perp}\right]\cdot \x\right)\right)},
\end{aligned}
\end{equation}
where the amplitude term $A(\x, \lambda)$ is approximately
\begin{align}
    A(\x, \lambda) \approx A\!\left(\x - s\!\left[\frac{\lambda\valpha_i}{\lambda_c} + \n_{\perp}\right]\!\right).
\end{align}
Similar to $M_0$, the metasurface $M_i$ imparts an additional deflection through a linear phase term:
\begin{align}
    M_i(\x, \lambda) \approx b_i(\lambda)
    \exp\!\left(j\frac{2\pi}{\lambda_c}\vbeta_i\cdot\x\right),
\end{align}
where $b_i(\lambda)$ represents the transmission efficiency of different wavelengths and $\vbeta_i$ is the deflection vector.

Assuming the eyepiece is an ideal thin lens with focal length $f$, its phase modulation is modeled as:
\begin{align}
    L_i(\x) = (\lVert\x-\x_i\rVert < r_i)\,
    \exp\!\left(-\frac{j\pi\lVert\x-\x_i\rVert^2}{\lambda f}\right),
\end{align}
where $r_i$ is the aperture radius and $\x_i$ the lens center. The photosensor is positioned at the back focal plane of the lens, and an optical filter $F_i(\lambda, \pol)$ modulates the transmitted amplitude according to wavelength $\lambda$ and polarization state $\pol$. The resulting intensity distribution on the sensor plane is
\begin{align}
    h_i(\x, \lambda, \n) = c(\lambda, \pol)\,
    h_i(\x - f\n_\perp, \lambda),
\end{align}
where
\begin{align}
    c(\lambda, \pol) = \big|a_1(\lambda)b_i(\lambda)F_i(\lambda, \pol)\big|^2.
\end{align}
The point spread function (PSF) of the $i$th optical channel at wavelength $\lambda$, $h_i(\x, \lambda)$, is given by
\begin{align}
    h_i(\x, \lambda) =
    \left|\widehat{A}\!\left(\frac{\x - \Delta\x_i(\lambda) - \x_i}{\lambda f}\right)\right|^2,
\end{align}
where $\widehat{A}(\cdot)$ denotes the Fourier transform of the entrance pupil $A(\x)$, and the wavelength-dependent shift is
\begin{align}
    \Delta\x_i(\lambda) = \frac{\lambda f}{\lambda_c}(\valpha_i + \vbeta_i).
    \label{eq:dispersion}
\end{align}
\Cref{eq:dispersion} reveals that by jointly engineering $\valpha_i$ and $\vbeta_i$, the PSF dispersion---that is, the spectral shift of the focal spot---can be precisely tuned or even eliminated (when $\valpha_i + \vbeta_i = 0$). As illustrated in \cref{fig:image-formation-model}, one channel can achieve an achromatic focus, while another can intentionally preserve controlled chromatic dispersion for spectral encoding.

\paragraph{Image Formation Model.}
When imaging a scene that is within the depth of field and has hyperspectral radiance $H(\x, \lambda)$, the captured image at the $i$th channel, named as the $i$th \textit{sub-image}, can be modeled as follows:
\begin{align}
\begin{split}
    I_i&(\x;j) \!\! \substack{\\}= \text{Gauss}(0,\sigma^2) + \\
    &G\cdot\text{Poisson}\left (t \int \eta(\lambda;j) \left( H(\x,\lambda) \odot h(\x, \lambda) \right) d\lambda\right).
\end{split}
\label{eq:cam}
\end{align}
where $G$ and $t$ denote the sensor gain and exposure time, $\eta(\lambda;j)$ is the photosensor’s spectral response to color $j\in\{R,G,B\}$, and $\odot$ represents spatial convolution over the wavelength dimension.

The proposed optical system enables simultaneous capture of achromatic and dispersive images of the scene. Furthermore, each sub-image can adopt an independent configuration—such as distinct exposure settings and polarization or spectral sensitivities—through the insertion of specific optical filters $F_{1:V}$. The resulting measurements are then jointly processed by computational algorithms for tasks such as hyperspectral reconstruction, HDR fusion, and polarization imaging.

\subsection{Postprocessing Algorithm}
\label{sec:post}

The sub-images $I_{1:V}$ can be directly processed by a standard image restoration network for hyperspectral reconstruction. In this work, we investigate two post-processing algorithms that build upon representative non-diffusion and diffusion-based architectures.

The first method largely follows adaptations of the Deep Wiener Deconvolution Network (DWDN)~\cite{liu2025metah2, dong2021dwdn} in hyerpsectral reconstruction~\cite{shi2024split,liu2025metah2}, which first perform Wiener deconvolution $I_{1:V}$ in the feature domain, followed by a multi-scale refinement through a feed-forward convolutional network.

The second builds upon Hazineh \etal~\cite{hazineh2024grayscale}, which employs a Denoising Diffusion Probabilistic Model (DDPM)~\cite{ho2020denoising}. In this framework, the DDPM is first trained to reconstruct a full hyperspectral irradiance datacube $H$ from $V$ input patches, $P_{1:V}$. Each patch $P_i$ is extracted from one sub-image $I_i$ and is spatially aligned with the regions corresponding to the other patches. 

During inference, the sub-images $I_{1:V}$ are partitioned into non-overlapping patches, denoted as $\{P_{1:V}^k\}$. For each index $k$, the $V$ patches correspond to the same spatial region across the sub-images. At each diffusion time step $t \in [0, T]$, the denoiser predicts a hyperspectral datacube $H^{k,t}$ for each patch index $k$. To enforce spatial consistency across the reconstructed hyperspectral datacubes $\{H^{k,t}\}$ for different indices $k$, the model further estimates a normalization factor $a^{k,t}$ by minimizing the discrepancy between the rendered measurements from $a^{k,t}H^{k,t}$ using \cref{eq:cam} and the measured patch $P_{1:V}^k$~\cite{hazineh2024grayscale}.

In our implementation, we augment this normalization process with an additional bias term, producing $a^{k,t}H^{k,t} + b^{k,t}$, and apply a scheduled learning rate to the discrepancy minimization. We empirically observe that both modifications improve reconstruction quality. Full implementation details are provided in the supplement.

\section{Experimental Results}

\subsection{Prototyping}
\label{secsec:prototype}

\begin{figure*}
    \centering
    \includegraphics[width=\linewidth]{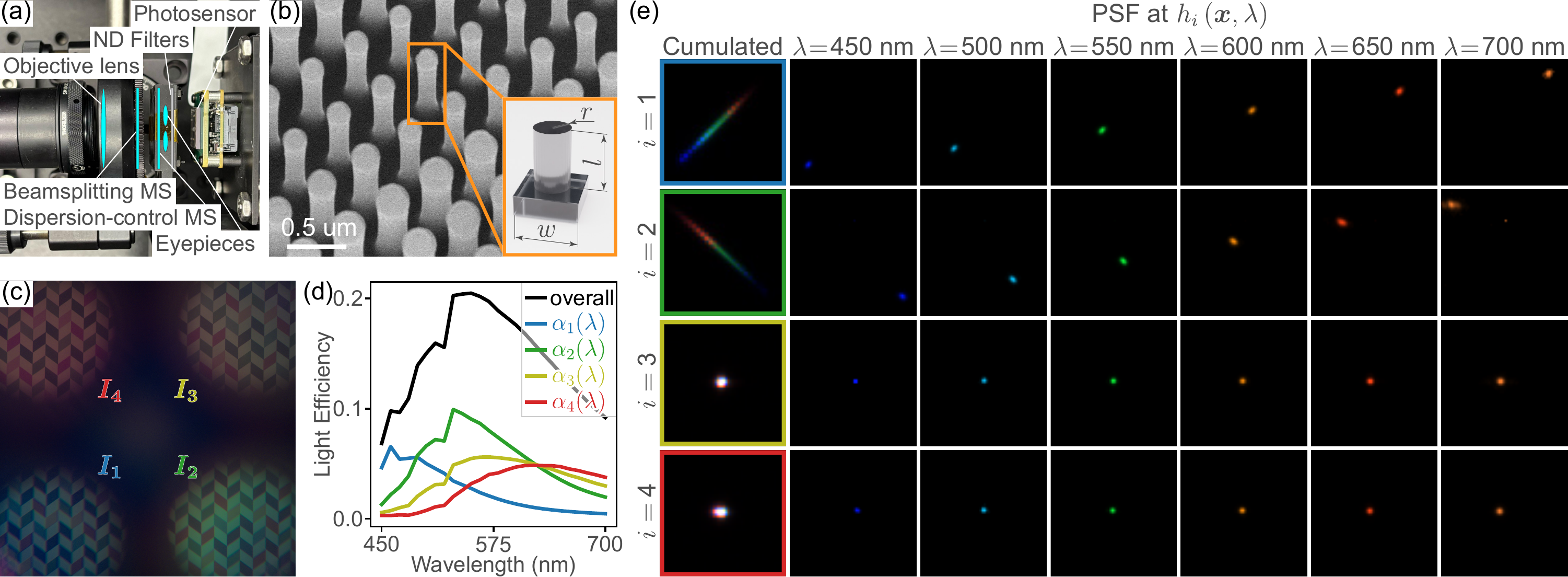}
    \caption{MetaSpectra+ prototype.
(a) Photograph of the experimental setup. The metasurfaces (MS), lenses, filters, and photosensors are mounted on a combination of commercial opto-mechanical components and custom 3D-printed holders. Additional implementation details are provided in the supplement. 
(b) Sample SEM image of the metasurfaces used in the system. The devices employ a SiN nanocylinder array for phase and dispersion control; the inset illustrates the nanocell parametrization. 
(c) Representative raw measurements. Sub-images $I_1$ and $I_2$ exhibit orthogonal dispersion, while $I_3$ and $I_4$ remain achromatic. A picture of the target is included in the supplement. Zoom in for fine details. 
(d) Calibrated spectral responses $\alpha_{1:4}(\lambda)$. The varied peak wavelengths are due to the different design wavelengths $\lambda_c$ for each sub-image. 
(e) Measured point spread functions across wavelengths. The PSFs for $I_1$ and $I_2$ show wavelength-dependent lateral shifts with consistent focus, whereas those for $I_3$ and $I_4$ remain spatially stationary, confirming their achromatic behavior.}
    \label{fig:experimental-prototype}
    \vspace{0.1in}
\end{figure*}

The prototype system is shown in \cref{fig:experimental-prototype}a. It comprises off-the-shelf lenses and image sensors, custom-fabricated metasurfaces, and a 3D-printed housing. The complete list of components is provided in the supplement. The prototype operates over a broad visible region from $450$~nm to $700$~nm.

\paragraph{Metasurface Design and Fabrication.}
The beamsplitting metasurface $M_0$ divides the incident beam into $V=4$ optical channels arranged in a $2\times2$ grid, each deflected by approximately $33^\circ$. The dispersion-control metasurfaces for the first two channels ($M_1, M_2$) introduce orthogonal dispersions in the sub-images $(I_1, I_2)$. In contrast, the remaining two metasurfaces ($M_3, M_4$) are designed to cancel the deflection introduced by $M_0$, forming achromatic sub-images $(I_3, I_4)$ on the photosensor. These functionalities are verified by the captured sub-images shown in \cref{fig:experimental-prototype}c and the measured PSFs $h_i(\x, \lambda)$ visualized in \cref{fig:experimental-prototype}e. The exact deflection parameters $\valpha_i$ and $\vbeta_i$ are provided in the supplement.

We observe that the diffraction efficiency $a_1(\lambda)$ in \cref{eq:deflect1} decreases as $\lambda$ deviates from the design wavelength $\lambda_c$. To ensure that the entire visible spectrum is effectively captured by at least one optical channel, we assign distinct design wavelengths across the four channels: $\lambda_{c,1:4} = \{450, 550, 600, 750\}$~nm. 

The beamsplitting metasurface (also serving as the entrance pupil) and the dispersion-control metasurfaces have diameters of $2$~mm and $4$~mm, respectively, with an interlayer spacing of $s=4$~mm. These geometric parameters are optimized through simulation to mitigate vignetting and minimize PSF spatial variation across both the FoV and the wavelength band.

The metasurfaces are designed and fabricated following the process described in Brookshire \textit{et~al.}~\cite{brookshire2024metahdr}. A metasurface is represented as a 2D array of nanocells of width $w=300$~nm. Each nanocell contains a uniform-height ($h=775$~nm) cylindrical nanopillar with radius $r$. \Cref{fig:experimental-prototype}b shows an SEM image of the metasurface sample and a visualization of the nanocell. Thus, a metasurface can be parameterized by a 2D matrix $r(w\boldsymbol{k}), \boldsymbol{k}\in\mathbb{Z}^2$ that indicates the radius of each cylindrical nanopillar. Given a target optical modulation profile $M_i(\x)$, the corresponding nanocell arrangement is obtained by querying a precomputed library $\{r_n \rightarrow M(r_n; \lambda)\}$ to identify the nanocylinder radius $r_n$ that produces the closest modulation response to the target profile at each discretized position $w\boldsymbol{k}$:
\begin{align}
r_i(w\boldsymbol{k}) = \arg\min_{r_n} \big|M_i(w\boldsymbol{k}) - M(r_n; \lambda_{c,i})\big|.
\end{align}
Detailed fabrication procedures for realizing $r_i(w\boldsymbol{k})$ are provided in the supplement.

\paragraph{Interleaving of Beamsplitting Metasurface.} The beamsplitting metasurface $M_0$ spatially interleaves four optical modulation profiles $M_{0,1:4}$. In this work, we adopt a random interleaving strategy, as defined in \cref{eq:interleave}. An alternative is to interleave the modulation profile in a regular $2\times2$ mosaic pattern; however, simulations show that such structured interleaving introduces strong higher-order diffraction artifacts, partially due to the large deflection angles required by each channel. In contrast, random interleaving suppresses these artifacts, at the cost of a reduction in light efficiency~\cite{brookshire2024metahdr}. More detailed analysis is provided in the supplement.


\paragraph{System Integration.}
In addition to the metasurfaces, the prototype incorporates a 400~mm-focal-length achromatic doublet as the objective lens and four 12~mm-focal-length, 4~mm-diameter achromatic doublets as eyepieces. The depth of field (DoF) of the prototype is approximately from 0.2~\text{m} to 0.7~\text{m}, which can be adjusted by changing the objective lens. The four optical channels form images $I_{1:4}$ on a shared global-shutter photosensor with an active area of $7.1~\text{mm} \times 7.1~\text{mm}$, which can be configured with either an RGB or monochrome sensor. We use RGB sensor throughout the experiments of this paper. Filter holders are placed between each eyepiece and the photosensor, allowing the insertion of optional polarizers or neutral-density filters to enable additional imaging functionalities. See the supplement for a detailed assembly procedure.


\paragraph{Calibration.} Our prototype requires two calibration steps: geometric alignment and the spectral response characterization. The geometric alignment determines a homography between each sub-image $I_i$ and a pre-determined, reference sub-image $I_j$. We estimate these homographies by imaging a planar calibration target, detecting feature correspondences across sub-images, and solving for the homography using a robust matching algorithm. 

The spectral response characterization measures the system's light efficiency $\alpha_i(\lambda)$ for each wavelength $\lambda$ of each optical channel $i$. It is collectively determined by the light efficiency of each optical element and the photosensor. We first record the total energy $E(\lambda)$ received by the photosensor from a single-wavelength point light source in the absence of the optical assembly. This can be approximately measured by summing over photosensor readouts $I(\x)$ over all pixels $\x$, $E(\lambda) = \sum_{\x} I(\x)$. Then, we keep the point source and photosensor fixed, install the optical assembly, repeat the measurement process, and calculate the summed energy over each sub-image, $E_i(\lambda) = \sum_{\x} I_i(\x)$. The spectral response of each sub-image is quantified by the ratio of these two measurements multiplied by the photosensors' spectral response $\eta(\lambda)$:
\begin{align}
    \alpha_i(\lambda) = \eta(\lambda) E_i(\lambda)/E(\lambda).
\end{align} 
The calibrated $\alpha_i(\lambda)$ is shown in \cref{fig:experimental-prototype}d. The peak of each sub-image's light efficiency approximately follows the designed wavelength $\lambda_{c, 1:4}$, with slight shifts due to the quantum efficiency of the photosensor. 

\paragraph{Training Details.} We trained both post-processing architectures described in Sec.~\ref{sec:post}---the DWDN and the DDPM---using synthetic data generated from the Harvard~\cite{chakrabarti2011statistics} and ICVL~\cite{arad2016sparse} datasets. Sub-images were rendered according to \cref{eq:cam} using PSFs synthesized by the D-Flat simulator~\cite{hazineh2022d} based on our optical design. All images are normalized to between 0 and 1 across all sub-images throughout our experiment. We set the noise level $\sigma$ randomly sampled from a uniform distribution from 0.001 to 0.01. Additional details are in the supplement.





\subsection{Snapshot Hyperspectral Reconstruction}
\label{secsec:hs}

First, we compare the proposed system against prior work on hyperspectral reconstruction, focusing specifically on snapshot approaches. We select several representative methods, including those using end-to-end–optimized optics~\cite{shi2024learned, shi2024split, arguello2021shift, baek2021single}, spectrum-from-defocus techniques~\cite{aydin2025spectrum, foley2025spectral}, and RGB-to-spectrum liftup approaches~\cite{cai2022mask, zhao2020hierarchical}. \Cref{tab:comparison} summarizes the specifications of the experimental prototypes reported in these works. Among all compared systems, MetaSpectra+ provides the largest number of simultaneously measured sub-images while also achieving the smallest total track length—a practical indicator of system compactness—enabled by the use of metasurfaces. 

\begin{table*}
\caption{System-level comparison of recent hyperspectral imagers.
These systems exhibit complementary strengths in F-number, field of view (FoV), and compactness (quantified by total track length, TTL). MetaSpectra+ achieves the smallest TTL, enabled by the use of metasurfaces in its hybrid design. It also delivers the highest reconstruction accuracy across all metrics on the KAUST dataset.  *~indicates systems can simultaneously capture an achromatic RGB image for structural guidance.}
\label{tab:comparison}
\centering
\resizebox{\textwidth}{!}{
\begin{tabular}{@{}l@{\hskip 0.04in}l@{\hskip 0.05in}c@{\hskip 0.05in}c@{\hskip 0.05in}c@{\hskip 0.1in}c@{\hskip 0.1in}c@{\hskip 0.1in}c@{\hskip 0.05in}c@{}}
\toprule
\multicolumn{2}{@{}l}{\multirow{2}{*}{Method}} &
\multirow{2}{*}{Venue} &
\multirow{2}{*}{Optics} &
\multicolumn{4}{c}{System Specification} &
Hyperspectral Reconstruction \\
\cmidrule(lr){5-8} \cmidrule(lr){9-9}
& & & & \# \hspace{-0.05in} sub-images & $f\hspace{-0.02in}/\hspace{-0.02in}\#$ & FoV ($^\circ$) & TTL (mm) & PSNR (dB) \hspace{-0.07in} $\uparrow$ \hspace{-0.04in} / \hspace{-0.04in} SSIM \hspace{-0.07in} $\uparrow$ \hspace{-0.04in} / \hspace{-0.04in} SAM \hspace{-0.07in} $\downarrow$ \\
\midrule

\multirow{8}{*}{\rotatebox{90}{Snapshot}} & \textbf{Ours (DWDN)} & \multirow{2}{*}{--} & \multirow{2}{*}{MS + Lens} & \multirow{2}{*}{$4^{\ast}$} & \multirow{2}{*}{6} & \multirow{2}{*}{12} & \multirow{2}{*}{17} & \underline{32.92} / \underline{\textbf{0.94}} / \underline{\textbf{0.17}} \\
 & \textbf{Ours (DDPM)} &  &  &  &  &  &  & \underline{\textbf{33.31}} / \underline{0.92} / 0.23 \\
 & 2-in-1 Cam~\cite{shi2024split} & SIGGRAPH\,\textquotesingle24 & DOE + Lens & $2^{\ast}$ & 5.8 & 24 & 50 & 31.14 / 0.86 / 0.24 \\
 & Array-HSI~\cite{shi2024learned} & SIGGRAPH\,Asia\,\textquotesingle24 & DOE + CFA & 4 & 6.7 & 12 & 20 & 27.44 / 0.89 / \underline{0.20}\\
 & SCCD~\cite{arguello2021shift} & Optica\,\textquotesingle21 & DOE + CCA & 1 & 16.7 & 30 & 50 & 26.78 / 0.81 / 0.36 \\

 & Baek \etal~\cite{baek2021single} & ICCV\,\textquotesingle21 & DOE & 1 & 16.7 & 46 & 50 & 26.68 / 0.74 / 0.39 \\

 & MST++~\cite{cai2022mst++} & CVPRW\,\textquotesingle22 & Lens & $1^{\ast}$ & -- & -- & -- & 21.85 / 0.68 / 0.32 \\
 
 & HRNet~\cite{zhao2020hierarchical} & CVPRW\,\textquotesingle20 & Lens & $1^{\ast}$ & -- & -- & -- & 23.03 / 0.76 / 0.31 \\
 
\midrule

\multirow{2}{*}{\rotatebox{90}{\makecell[c]{Multi-\\[-0.8ex]shot}}} &  
Foley \etal~\cite{foley2025spectral} & ICCP\,\textquotesingle25 & Lens + CFA & 1-5 & 5.8 & 9.8 & 140 & 20.40 / 0.68 / 0.29\\

& SfD~\cite{aydin2025spectrum} & arXiv\,\textquotesingle25 & Lens & 5 & 2 & 10 & 44.5 & 27.54 / 0.82 / 0.40\\

\bottomrule
\end{tabular}}
\vspace{0.15in}
\end{table*}

\begin{figure*}
    \centering
    \includegraphics[width=\linewidth]{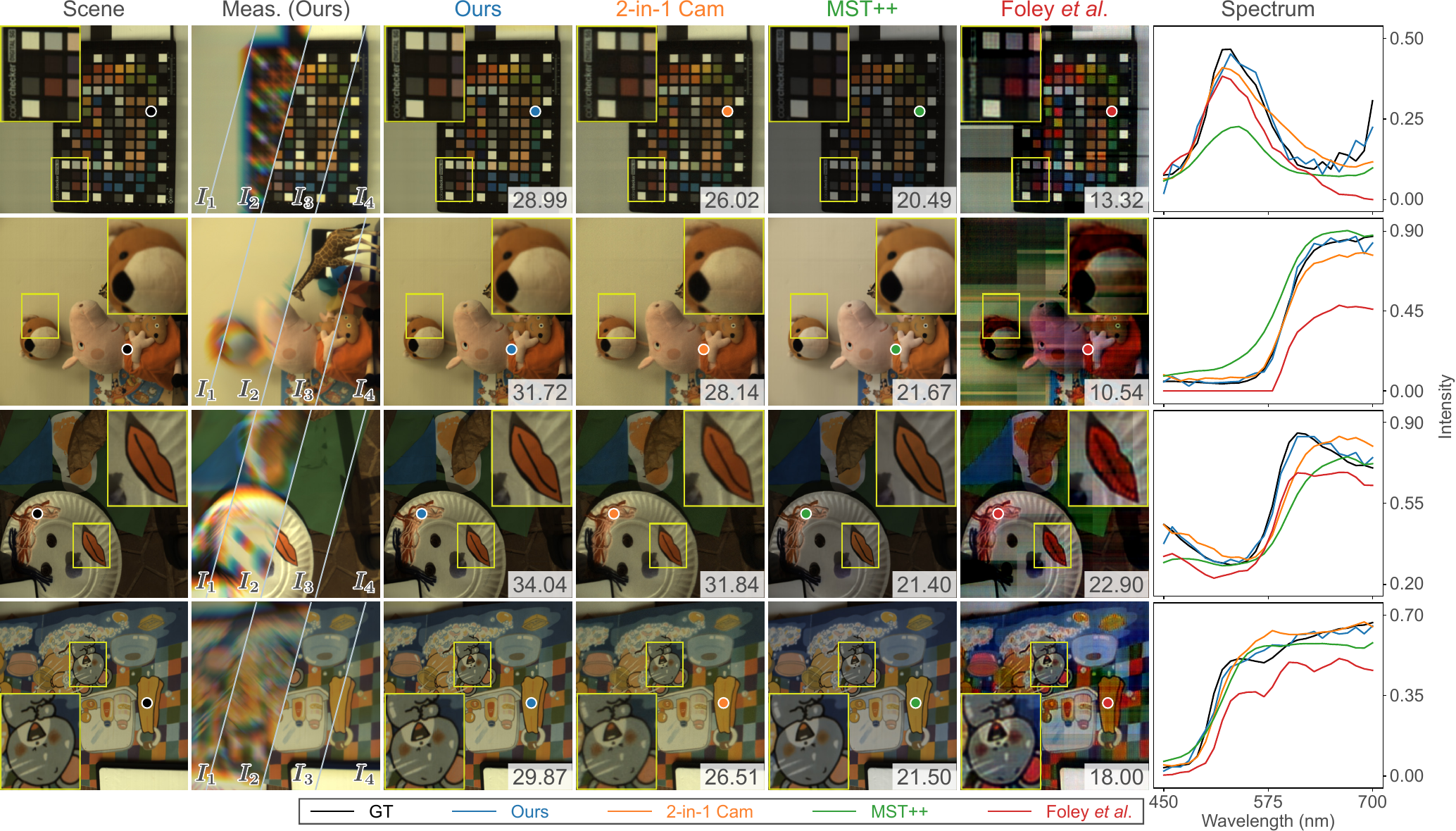}
    \caption{Sample hyperspectral reconstruction results on the KAUST dataset. MetaSpectra+ produces the highest structural fidelity and spectral accuracy among all compared methods. See enlarged insets for details.  The inset numbers are PSNR (dB) for hyperspectral reconstructions.}
    \label{fig:HS-experiment}
\end{figure*}

\paragraph{Simulation Comparison.} We evaluate the hyperspectral reconstruction accuracy of all methods on the KAUST dataset~\cite{li2021multispectral}, which is not used for training by any of the compared approaches. For each method, we generate the raw measurements using the provided renderer. We note that 450–700~nm is the spectral range shared by the majority of competing methods, and our main quantitative results are therefore reported on this common band. For methods whose spectral coverage or resolution cannot be aligned to this interval, the result is generated under the original spectral grid. \Cref{tab:comparison} reports the reconstruction accuracy on the KAUST dataset. Our method achieves the best performance on all metrics, demonstrating the state-of-the-art snapshot hyperspectral imaging capability of MetaSpectra+. \Cref{fig:HS-experiment} shows representative reconstructed RGB images and spectra from the KAUST dataset across different methods. MetaSpectra+ preserves finer spatial structures and achieves superior spectral fidelity both qualitatively and quantitatively. Additional results are in the supplement.

\begin{figure*}
    \centering
    \includegraphics[width=\linewidth]{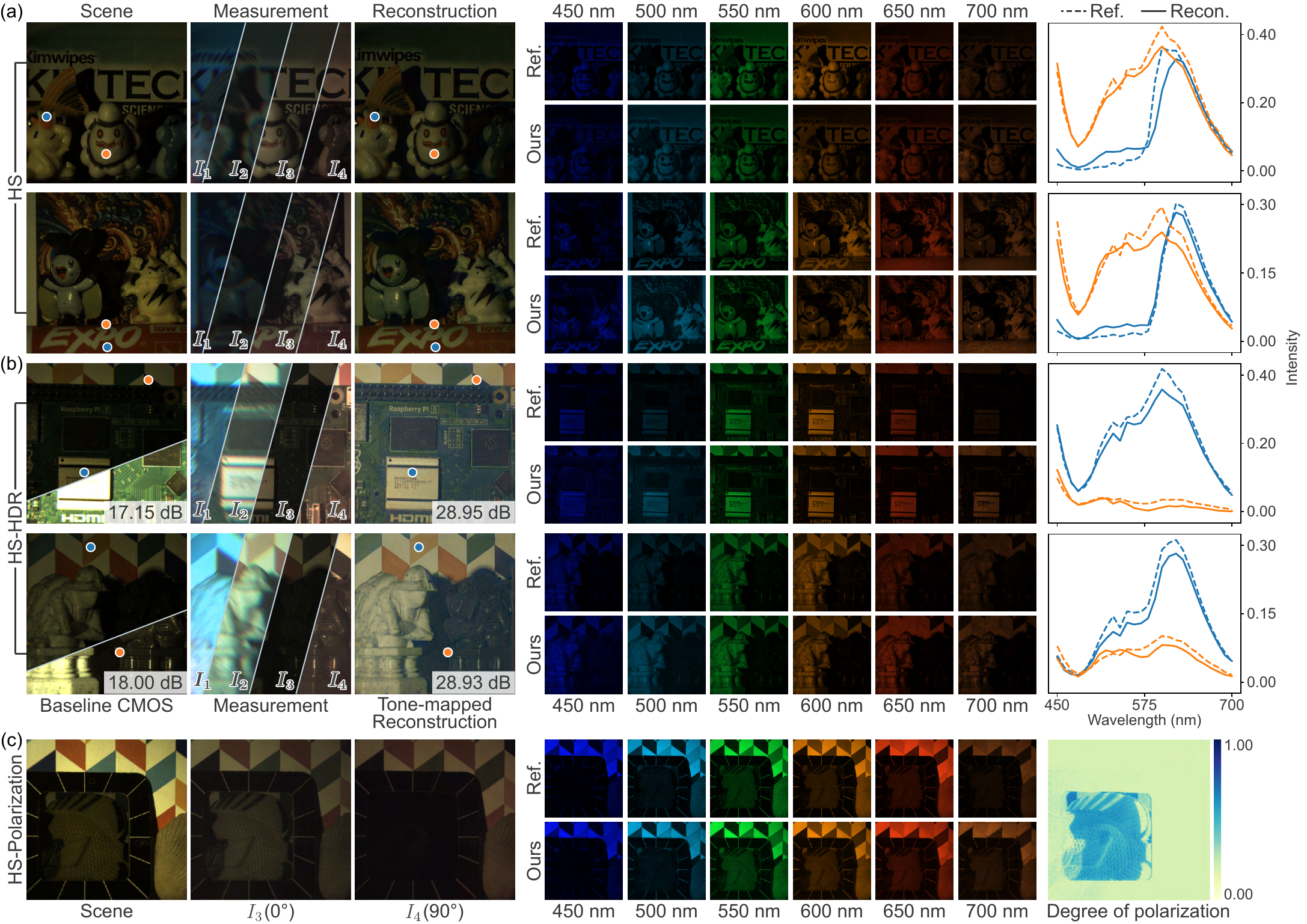}
    \caption{Sample real-world results of MetaSpectra+. (a) Hyperspectral imaging only. (b) HDR + hyperspectral imaging. Inset numbers represent the dynamic range (dB) of the picture. Compared to low-dynamic range (LDR) images recorded with CMOS cameras, the reconstructed HDR images from MetaSpectra+ demonstrate increases of 11~\text{dB} in dynamic range, and preserve both dark and bright scene details. Zoom in for finer structures. Inset numbers represent the dynamic range (dB) of the scene. (c) Polarization + hyperspectral imaging. The sample scene includes a $0^\circ$ linear polarizer, which appears dark in $I_4$ since this sub-image measures the $90^\circ$ polarization component.}
    \label{fig:real-HS-experiment}
\end{figure*}

\paragraph{Real-World Experiments.}
We further evaluate the MetaSpectra+ prototype on eight real-world scenes with different colors, textures, and shapes set up in a lab environment. Among them, three scenes contain only front-parallel textured patterns, while the remaining five include common objects with diverse 3D geometries. The reference hyperspectral cubes are obtained by capturing a sequence of images of the fixed scene with narrow bandpass filters that collectively cover the entire visible spectrum. To mitigate the systematic bias between the simulation and real measurements,  we fine-tune the reconstruction model using only the front-parallel scenes, and use the remaining scenes as a validation set for subsequent analysis. Textures in the scenes for fine-tuning do not appear in the validation scenes.
\Cref{fig:real-HS-experiment}a presents results on the five validation scenes, where the recovered hyperspectral datacubes exhibit high structural fidelity and spectral accuracy. More information about the scenes are in the supplement.

\subsection{Real-World Snapshot Hyperspectral+ Imaging}

\paragraph{HDR + Hyperspectral.}
We insert a neutral density (ND) filter with 0.3 optical density (OD) for sub-images $I_{1:3}$ and a 0.9 OD ND filter for $I_4$ to adjust their effective exposures. The filters are placed after the eyepiece (\cref{fig:image-formation-model}). Since $I_3$ and $I_4$ share similar spectral responses, they effectively form an exposure bracket with an approximate power ratio of 4, enabling up to 12.04~dB additional dynamic range beyond the native sensor capability. We fuse $I_3$ and $I_4$ using the classic method of Debevec and Malik~\cite{debevec2023recovering} to obtain an HDR estimate $I_{\text{HDR}}$, and then feed $I_{\text{HDR}}$, $I_1$, and $I_2$ into the same postprocessing module described in Sec.~\ref{secsec:hs} to generate the hyperspectral datacube. \Cref{fig:real-HS-experiment}b shows sample results of MetaSpectra+ on real-world HDR scenes. The reconstructed, tonemapped RGB images retain details in both dark and bright regions, and the recovered spectra remain accurate. Additional results are in the supplement.

\paragraph{Polarization + Hyperspectral.}
We place a $0^\circ$ and a $90^\circ$ linear polarizer in front of $I_3$ and $I_4$, respectively, while leaving $I_{1:2}$ unfiltered. In this configuration, the achromatic channels $I_3$ and $I_4$ measure the $0^\circ$ and $90^\circ$ polarization components of the scene, whereas $I_1$ and $I_2$ capture the dispersed images independent of polarization. We generate the hyperspectral datacube using $I_3 + I_4$, $I_1$, and $I_2$ with the same postprocessing pipeline described earlier, and compute the horizontal–vertical degree of linear polarization (DoLP$_\text{HV}$) as $\text{DoLP}_\text{HV} = |I_3 - I_4| / |I_3 + I_4|$. \Cref{fig:real-HS-experiment}c shows sample results of joint polarization and hyperspectral imaging. Additional examples are in the supplement. \\

\noindent \textbf{Acknowledgement.} The metasurfaces in this work were fabricated by SNOChip Inc. through their custom metasurface fabrication service according to the authors’ specifications. We thank Professor Jian Jin for helpful discussions. This work was supported by the National Science Foundation (NSF) under Grant No. CCF-2431505.
\newpage

{
    \small
    \bibliographystyle{ieeenat_fullname}
    \bibliography{main}
}

\clearpage
\setcounter{page}{1}
\maketitlesupplementary


\begin{table}[b]
  \caption{Hyperparameters of DDPM.}
  \label{tab:ddpm_hyperparam}
  \centering
  \begin{tabular}{@{}lc@{}}
    \toprule
    Item & Value \\
    \midrule
    Patch size & $128 \times 128$ \\
    Batch size & 64 \\
    Input channels & 38 \\
    Output channels & 26 \\
    Time steps & 1000 \\
    Time embedding dimension & 1024 \\
    U-Net block channels & [64, 128, 256, 512, 1024] \\
    Use attention & All blocks \\
    Attention head dimension & 32 \\
    Classifier-free guidance & False \\
    $\beta$ schedule & Linear, 1e-4 $\rightarrow$ 2e-2 \\
    Learning rate schedule & Cosine, 1e-4 $\rightarrow$ 1e-6 \\
    Min-SNR weighting & True \\
    $K_{\text{min}}$-SNR & 5.0 \\
    Loss type & L1 (on noise) \\
    Optimizer & AdamW \\
    Epochs & 15000 \\
    \bottomrule
  \end{tabular}
\end{table}

\section{Implementation Details of the DDPM}

The DDPM used for hyperspectral reconstruction closely follows the architecture of Hazineh~\etal~\cite{hazineh2024grayscale}. During training, we randomly crop four uniform-sized patches—one from each sub-image $I_{1:V}$—and use them as the model input. At inference time, we partition the sub-images into $L$ non-overlapping patches $\{P_{1:V}^k\}_{k=1}^L$ of the same size. At each diffusion step $t \in [0, T]$, the DDPM employs a U-Net to estimate a hyperspectral datacube corresponding to each group of four patches with index $k$:
\begin{equation}
    \epsilon^{k,t} = \text{U-Net}\left( \state^{k,t}, \{P_{1:V}^{k}\} \right),
    \label{eq:est_noise}
\end{equation}
\begin{equation}
    H^{k,t} = \frac{\state^{k,t} - \sqrt{1-\upsilon^t} \epsilon^{k,t}}{\sqrt{1-\upsilon^t}},
    \label{eq:denoise}
\end{equation}
where $\state^{k,t}$ denotes the state of the $k$--th set of patches at time step $t$, $\epsilon^{k,t}$ is the corresponding estimated noise, and $\upsilon^t$ is the cumulative product of noise-retention factors up to time step $t$. The predicted datacubes $\{H^{k,t}\}_{k=1}^L$ are stitched together to form the full-scene hyperspectral estimate:
\begin{align}
    H^t(\x) = a^{k,t} H^{k,t}(\x) + b^{k,t}, \quad \text{where }  P_{1:V}^{k} \ni \x.
    \label{eq:stitch}
\end{align}
The per-patch scale and offset parameters $(a^{k,t}, b^{k,t})$ ensure consistency across neighboring patches and are obtained with the guidance of measured sub-images:
\begin{equation}
    \tilde{a}^{k,t}, \tilde{b}^{k,t}
    =
    \underset{\{a^{k,t}, b^{k,t}\}}{\arg\min}
    \left\|
    \hat{I}_{1:V}(H^t) - I_{1:V}
    \right\|^2,
    \label{eq:loss-ab}
\end{equation}
where $\hat{I}_{1:V}$ denotes the re-rendered sub-images generated from the predicted hyperspectral datacube $H^t$ via \cref{eq:cam}. Finally, the latent state $\state^{k,t}$ is updated via normalized gradient descent:
\begin{align}
    \hat{\state}^{k,t}
    =
    \state^{k,t}
    -
    \gamma(t)
    \frac{
        \nabla_{\state^{k,t}} \mathcal{L}
    }{
        \|\nabla_{\state^{k,t}} \mathcal{L}\|
    },
    \label{eq:iterative}
\end{align}
where $\mathcal{L}$ is the loss in \cref{eq:loss-ab} evaluated with the converged $(\tilde{a}^{k,t}, \tilde{b}^{k,t})$. We use a decaying step size $\gamma(t) = \sqrt{t/T}$, which we found empirically to improve reconstruction stability and accuracy. This guidance process (\cref{eq:denoise,eq:stitch,eq:loss-ab,eq:iterative}) can be done iteratively to refine scale and offset parameters. The updated state $\hat{\state}^{k,t}$ is then used for the next time step.

We train the DDPM on an NVIDIA GeForce RTX~5090 GPU with 32~GB of memory. The sampling process uses 50 denoising diffusion steps and 20 guidance iterations. \Cref{tab:ddpm_hyperparam} lists all training hyperparameters used in our implementation. 

\section{Metasurface Design and Fabrication}
\subsection{Fabrication Procedure}

The fabrication process for our metasurfaces is identical to the process described in Brookshire~\etal~\cite{brookshire2024metahdr}, which can be performed in a typical university cleanroom. It begins with coating a fused silica substrate with a $775$~nm thick silicon nitride layer deposited via plasma-enhanced chemical vapor deposition. Then, it spin coats a $200$~nm ARP6200 resist layer onto the surface, followed by depositing a 20-nm aluminum conductive layer. The designed pattern is printed onto the sample using e-beam lithography with an electron dose of $700$~$\text{uC/cm}^2$.  After exposure, the Aluminium layer is removed, followed by a $90$~s development step. Finally, we deposit a $30$~nm $\mathrm{Al_2O_3}$ layer and perform lift-off, forming a hard mask for subsequent etching. The designed nanostructures are transferred into the silicon nitride layer using plasma dry etching.

In addition to university cleanrooms, there are also several companies across different countries that offer custom metasurface fabrication services, ranging from single-sample prototyping to wafer-level production (e.g., $\sim100$ devices per wafer), with costs spanning from several thousand to hundreds of thousands of USD.
\subsection{Details of Metasurface Design}

\begin{figure}
    \centering
    \includegraphics[width=1.0\linewidth]{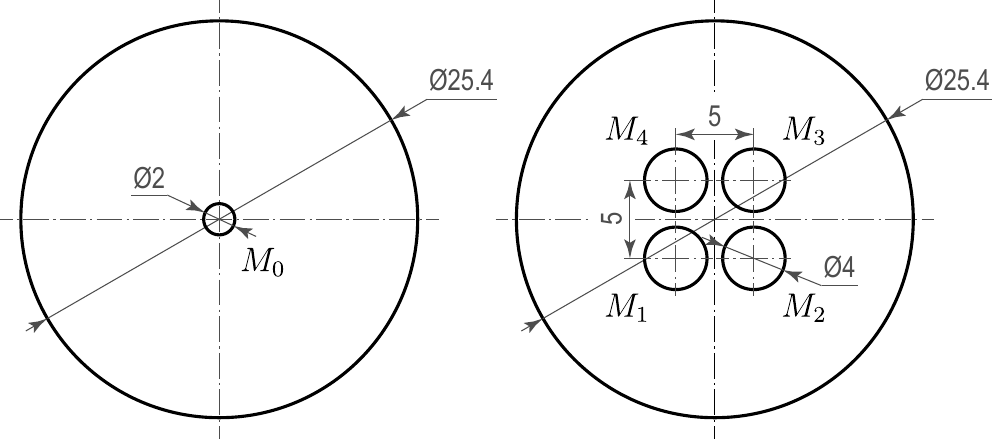}
    \caption{Schematics of the beamsplitting metasurface $M_0$ (left) and the dispersion-control metasurfaces $M_{1:4}$ (right). Units: mm.}
    \label{fig:ms_details}
\end{figure}

\paragraph{Deflection Vectors.} As shown in \cref{fig:ms_details}, the beamsplitting metasurface and the dispersion-control metasurfaces have clear apertures of $2$~mm and $4$~mm, respectively, separated by $s=4$~mm. The dispersion-control metasurfaces $M_{1:4}$ are arranged in a $2\times 2$ configuration with a center-to-center spacing of $5$~mm. The designed deflection vectors, $\valpha_i$ and $\vbeta_i$, are set to:
\begin{align}
\begin{split}
    &\valpha_i=\frac{\lambda_{c,i}}{\lambda_{c,2}}\left[ (-1)^{\lfloor i/2\rfloor+1}, (-1)^{\lfloor (i-1)/2\rfloor+1} \right]^{\top} \cdot \valpha_2,\\
    &\vbeta_i= -\valpha_i+
                \begin{cases}
                    \left[0.017,(-1)^{i-1}0.017\right]^{\top} &i=1,2\\ 
                    \boldsymbol{0}, &i=3,4
                \end{cases}, \\
    &\valpha_2 = [0.385,-0.385]^{\top}.
\end{split}
\end{align}
These parameters are selected to ensure the front-parallel incident beam with wavelength 550~nm to be deflected to the center of each dispersion-control metasurface. The dispersion of channels 3 and 4 is eliminated as $\valpha_i = \vbeta_i, i=3,4$, while channels 1 and 2 retain a small amount of dispersion whose amount is decided by simulation. 

\begin{figure}
    \centering
    \includegraphics[width=1.0\linewidth]{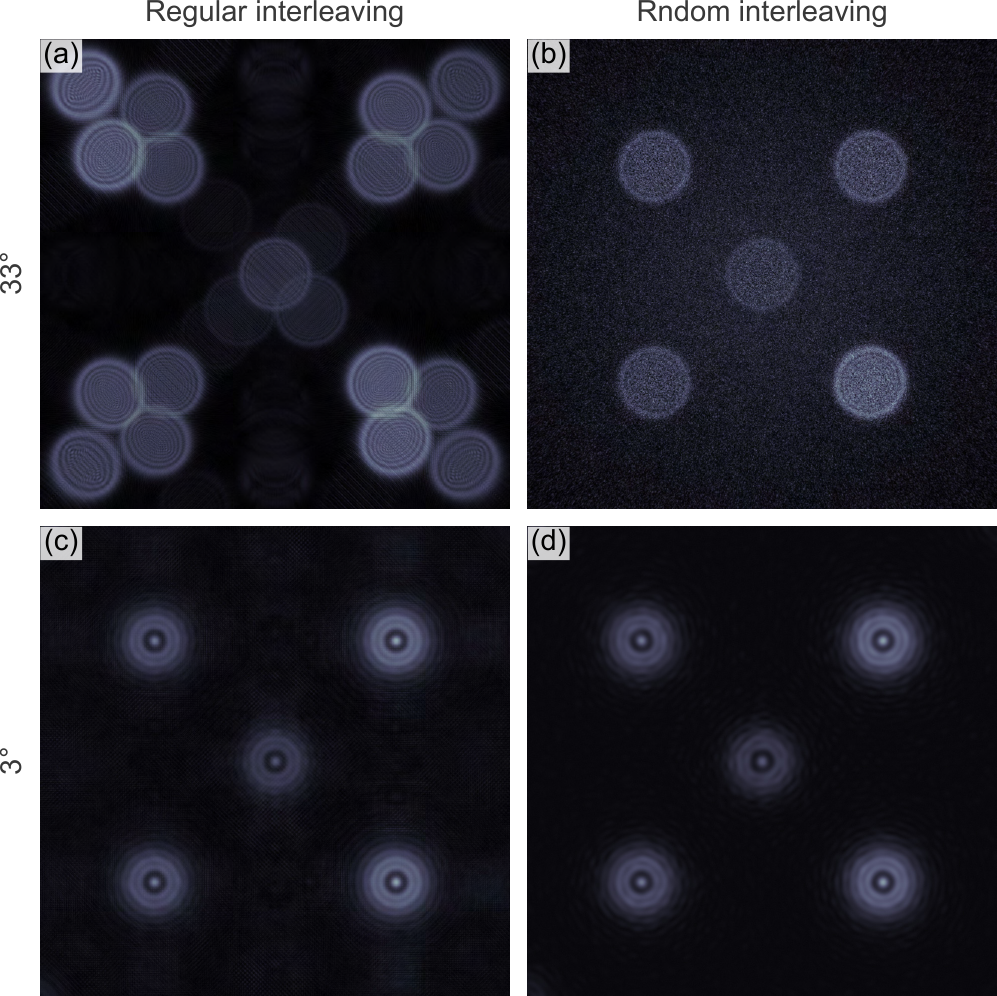}
    \caption{Simulated wavefront amplitude incident on the dispersion-control metasurface when using (a,c) regular interleaving and (b,d) random interleaving in the beamsplitting metasurface $M_0$ at different deflection angles. Random interleaving suppresses higher-order diffraction artifacts at the cost of increased background residual light. The central bright spot arises from zeroth-order diffraction when the system is operating at non-designed wavelength.}
    \label{fig:interleaving}
\end{figure}


\paragraph{Interleaving Strategy.} Regular interleaving, which periodically arranges multiple phase patterns on a uniform nanocell grid, has been widely used in metasurface design to realize multi-functionality~\cite{guo2019compact, khorasaninejad2016multispectral}. However, this strategy becomes problematic when the uninterleaved phase profile corresponds to large deflection angles. To illustrate this effect, consider regularly interleaving the four deflection patterns $M_{0,1:4}$ in a $2\times 2$ mosaic pattern:
\begin{align}
    M_0(\x;\lambda)=\sum_{i=1}^4 M_{0,i}(\x,\lambda) \ \text{comb}\left(\frac{\x-\Delta \x_i}{2w}\right),
\end{align}
where $\text{comb}(\cdot)$ is a 2D Dirac comb function and $\Delta \x_i$ controls the sampling position
\begin{align}
\Delta \x_i=
    \begin{cases}
        (0, 0) \quad & i=1\\
        (w, 0) \quad & i=2\\
        (w, w) \quad & i=3\\
        (0, w) \quad & i=4
    \end{cases}.
\end{align}
\Cref{fig:interleaving}a shows the simulated field distributions of $550$~nm at the plane immediately before the dispersion-control metasurface under different deflection angles and interleaving strategies. It clearly shows that regular interleaving leads to residual diffraction orders when the deflection angle becomes large, which will become repeated patterns on the image plane.  

Random interleaving avoids these periodic replicas by replacing the structured sampling with an irregular arrangement, but it also introduces a weak background due to random mixing of the four patterns~\cite{brookshire2024metahdr}. In our optical design, this trade-off makes random interleaving safer for multiplexing under large deflection angles. This can be evidenced in \cref{fig:interleaving}b, which shows four clear deflection patterns without residual diffraction orders under various deflection angles using random interleaving.

\begin{table*}[t]
\caption{List of parts.}
\label{tab:list}
\centering
\begin{tabular}{@{}l@{\hskip 0.106in}c@{\hskip 0.106in}c@{\hskip 0.106in}c@{\hskip 0.106in}c@{}}
\toprule
No. & Item & Vendor \& Stock Number & Quantity & Description \\
\midrule
1 & Objective lens & Thorlabs, AC254-400-A & 1 & $\diameter$ 25.4 mm, 400~mm focal length, achromatic \\
2 & Beam-splitting metasurface & SNOChip & 1 & $\diameter$ 2~mm, 1-inch substrate \\
3 & Dispersion-control metasurface & SNOChip & 1 & $\diameter$ 4~mm, 2$\times$2, 1-inch substrate,  \\
4 & Eyepiece & Edmund Optics, 63-704 & 4 & $\diameter$ 4~mm, 4~mm focal length, achromatic\\
5 & Neutral density (ND) filter & Edmund Optics, 62-662 & 1 & $\diameter$ 50 mm, optical density (OD) 0.3 \\
6 & Neutral density (ND) filter & Edmund Optics, 62-665 & 1 & $\diameter$ 50 mm, optical density (OD) 0.9 \\
7 & Linear Polarizer & Edmund Optics, 26-919 & 1 & 50~mm$\times$50~mm$\times$0.11~mm \\
8 & Photosensor & Basler, dmA3536-9gc & 1 & $3536\times 3536$ resolution, 2~$\mu$m pixel pitch, RGB \\
\bottomrule
\end{tabular}
\end{table*}

\section{Prototype}

\begin{figure}
    \centering
    \includegraphics[width=0.82\linewidth]{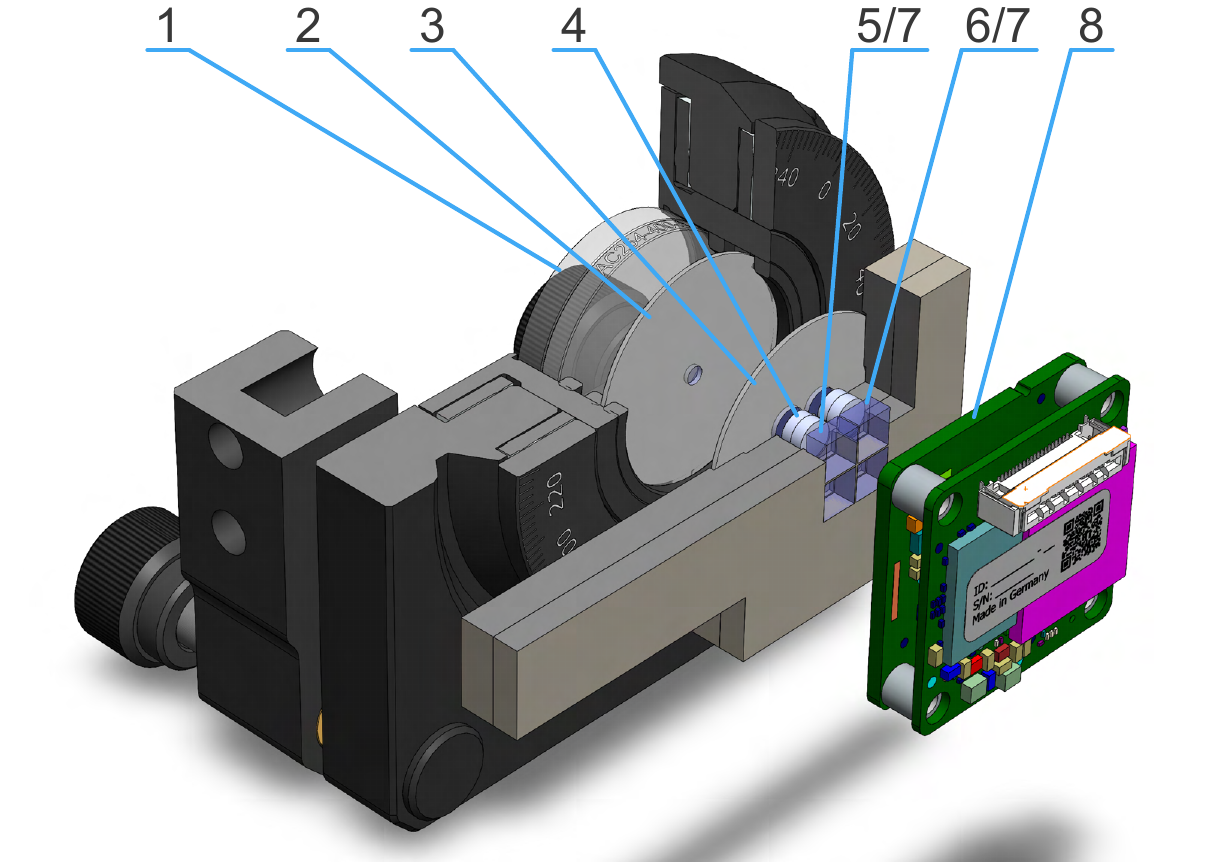}
    \caption{CAD model of the assembly. Numbers correspond to the items in the list of parts in \cref{tab:list}.}
    \label{fig:cad}
\end{figure}

\paragraph{Assembling Instruction.} The prototype is constructed from a combination of off-the-shelf components and custom 3D-printed parts. A full list of components is provided in \cref{tab:list} and the CAD model of the assembly is visualized in \cref{fig:cad}. The objective lens and the beam-splitting metasurface are mounted inside a rotation mount. The remaining optical elements are secured in a 3D-printed holder, 
where we design a circular recessed pocket and four holes to fit and align the dispersion-control metasurface and subsequent lenses. We cut off-the-shelf neutral density (ND) filters and polarizers to fit the filter holders using a water-jet blade. The 3D printed holder and photosensor are mounted on a three-axis translation stage to allow fine positional adjustment during installation.


\section{Additional Information}

\begin{figure}
    \centering
    \includegraphics[width=0.45\linewidth]{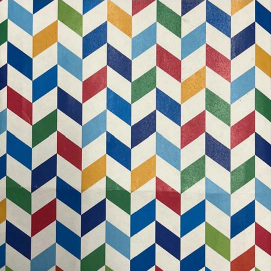}
    \caption{Picture of the target being measured in \cref{fig:experimental-prototype}c.}
    \label{fig:target_pic}
\end{figure}

\begin{figure}[h]
    \centering
    \includegraphics[width=\linewidth]{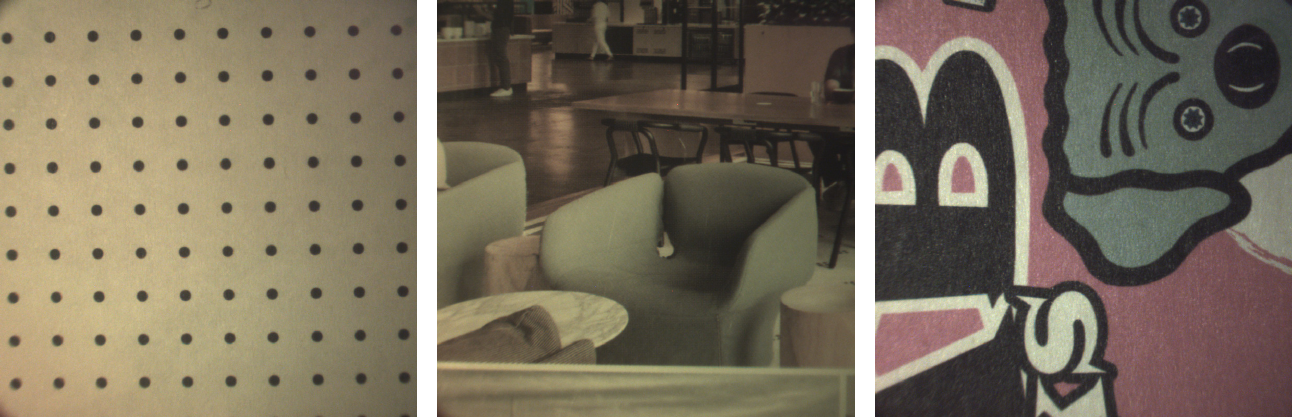}
    \caption{Sub-images $I_3$ of the real-world scenes for fine-tuning.}
    \label{fig:finetune_images}
\end{figure}

\Cref{fig:target_pic} shows the object being measured in \cref{fig:experimental-prototype}c. It is a printed, multi-colored pattern mounted in a front-parallel orientation.

The three scenes that only contain front-parallel textured planes used exclusively for fine-tuning the computational model are shown in \cref{fig:finetune_images}. Textures appearing in the fine-tuning scenes do not appear in the testing scenes to avoid information leakage.

\begin{figure*}[ht]
    \centering
    \includegraphics[width=\linewidth]{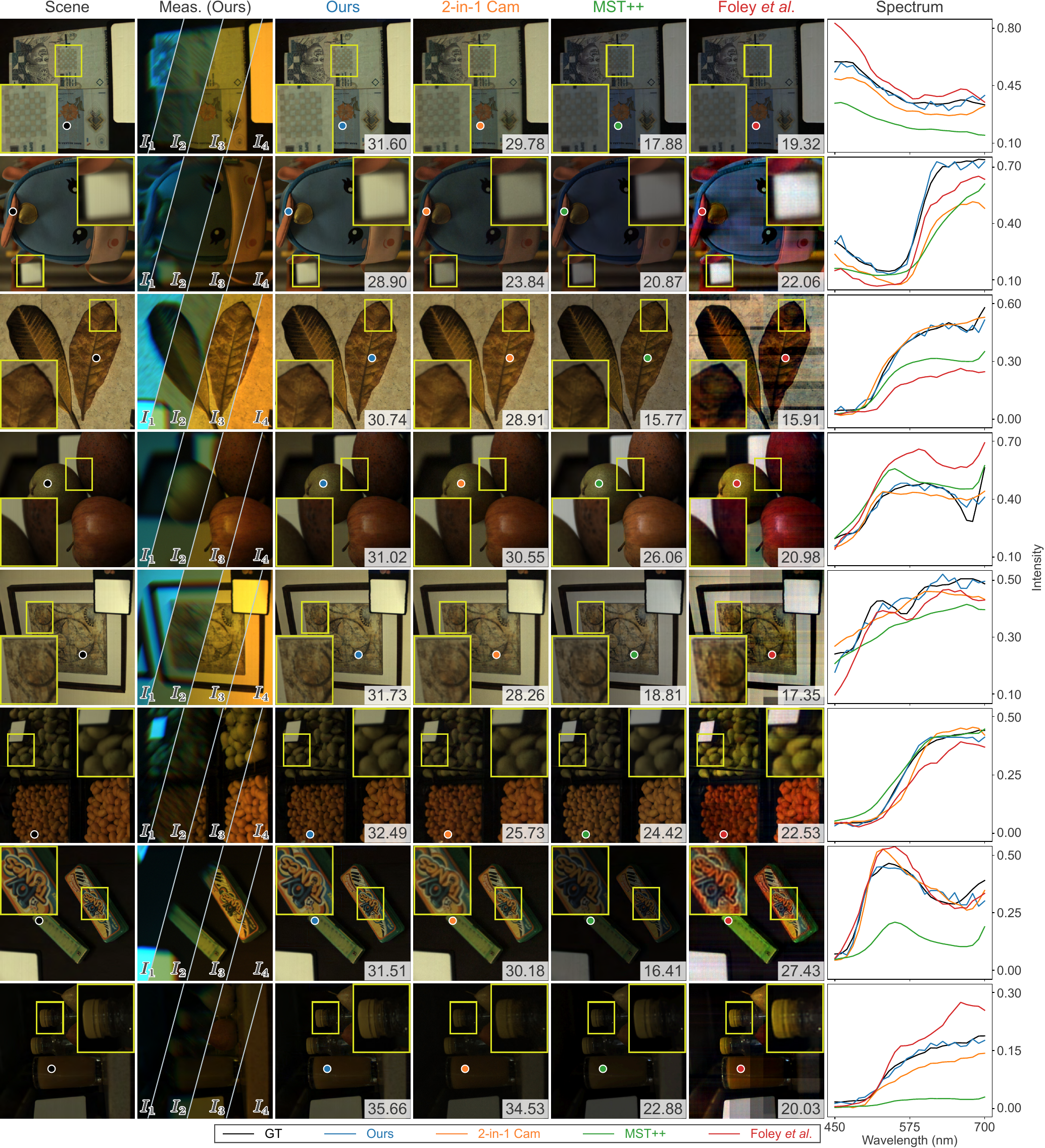}
    \caption{Additional sample hyperspectral reconstruction results on the KAUST dataset. The inset numbers are PSNR (dB) for hyperspectral reconstructions.}
    \label{fig:HS-experiment_suppl}
\end{figure*}

\begin{figure*}[ht]
    \centering
    \includegraphics[width=\linewidth]{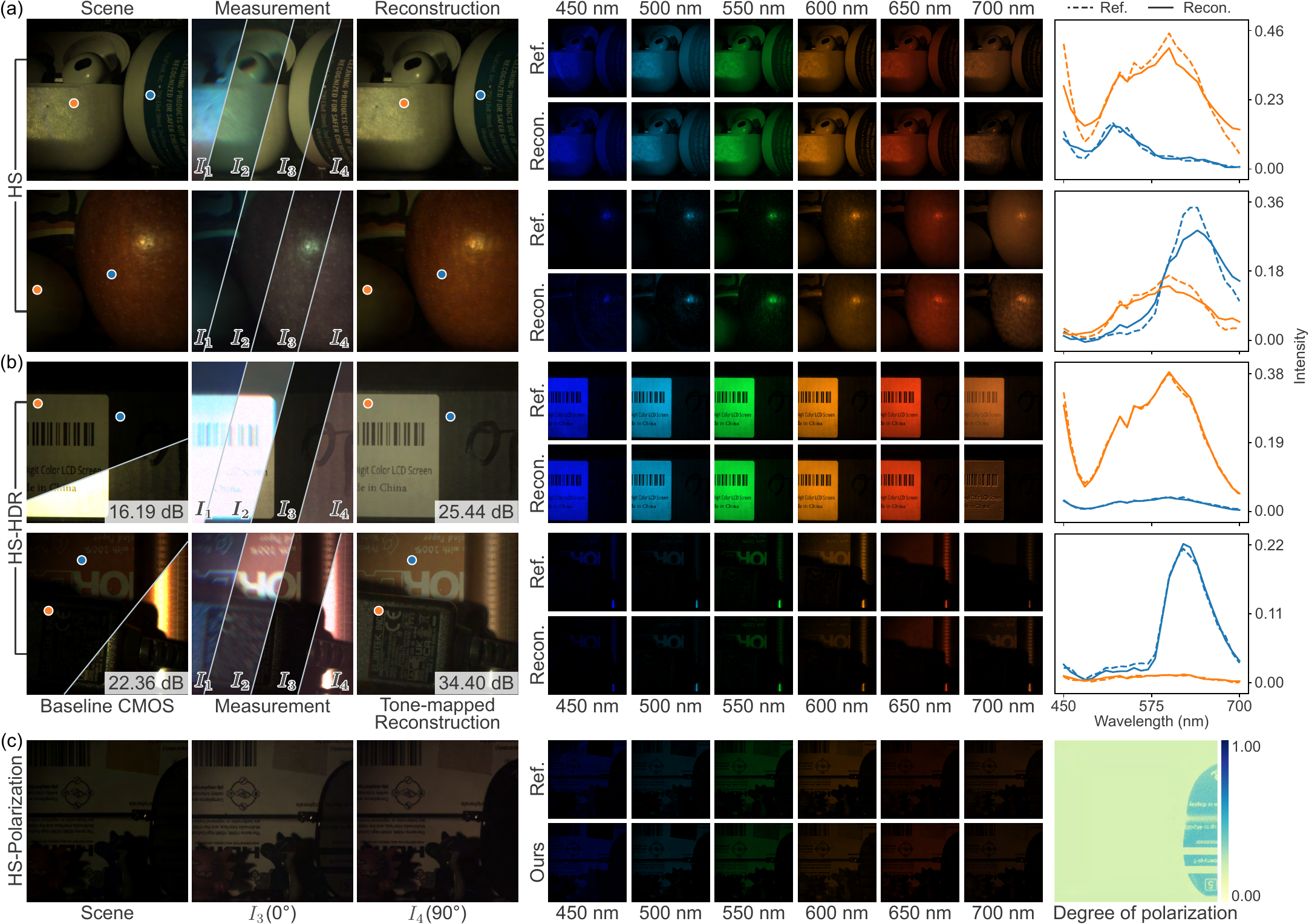}
    \caption{Additional sample real-world results of MetaSpectra+. Inset numbers in (b) represent the dynamic range (dB) of the picture.}
    \label{fig:real-HS-experiment_suppl}
\end{figure*}

\section{Additional Results}
\paragraph{Snapshot Hyperspectral Imaging.} \Cref{fig:HS-experiment_suppl} shows additional hyperspectral reconstruction results using synthetic data generated from the KAUST dataset. MetaSpectra+ consistently outperforms previous hyperspectral imaging solutions in terms of both metrics and visual quality. We also validate the proposed system's performance on additional real-world experiments. 
\Cref{fig:real-HS-experiment_suppl}a demonstrates low reconstruction errors and faithful spectral recovery across real-captured, diverse scenes, demonstrating the robustness and effectiveness of our system.


\paragraph{Snapshot Hyperspectral+ Imaging.} \Cref{fig:real-HS-experiment_suppl}b-c provides additional real-world experimental results for HDR + hyperspectral and polarization + hyperspectral imaging.


\section{Discussion}
\paragraph{Competing Approaches.} Besides the proposed solution, we identify three alternative approaches to realize the functionality of MetaSpectra+. Here, we discuss the advantages and disadvantages of each.

\begin{enumerate}
    \item Diffraction gratings can provide multi-beam splitting, and industrial off-the-shelf components are relatively cost-effective. However, implementing the specialized optical functionalities required in this work would necessitate custom-fabricated gratings, leading to costs comparable to metasurfaces.
    \item Diffractive optical elements (DOEs) enable custom wavefront shaping and typically offer lower cost and larger aperture sizes than metasurfaces. However, realizing the multi-beam-splitting behavior of this work within a single DOE is challenging.
    \item Refractive optics are generally the most cost-effective option; however, achieving comparable functionality using off-the-shelf components would significantly increase the system form factor, as the required multifunctionality would rely on cascaded beamsplitters.
\end{enumerate}

\paragraph{Limitations.} Due to a reduced diffraction efficiency caused by the random interleaving in the beam-splitting metasurface, an increased integration time is typically required. Consequently, our current prototype operates at up to 10 FPS, posing challenges for direct deployment in high-speed video applications. However, this issue can potentially be mitigated by optimizing the metasurface design, such as decreasing the beam deflection angle, or using materials with higher refractive indices, such as gallium nitride and titanium dioxide.

\end{document}